# wav2vec and its current potential to Automatic Speech Recognition in German for the usage in Digital History

## A comparative assessment of available ASR-technologies for the use in cultural heritage contexts

Michael Fleck, University of Graz (michael.fleck@uni-graz.at)

Wolfgang Göderle, University of Graz (wolfgang.goederle@uni-graz.at)

**Abstract**

In this case study we trained and published a state-of-the-art open-source model for Automatic Speech Recognition (ASR) for German to evaluate the current potential of this technology for the use in the larger context of Digital Humanities and cultural heritage indexation. Along with this paper we publish our wav2vec2 based speech to text model while we evaluate its performance on a corpus of historical recordings we assembled compared against commercial cloud-based and proprietary services. While our model achieves moderate results, we see that proprietary cloud services fare significantly better. As our results show, recognition rates over 90 percent can currently be achieved, however, these numbers drop quickly once the recordings feature limited audio quality or use of non-every day or outworn language. A big issue is the high variety of different dialects and accents in the German language. Nevertheless, this paper highlights that the currently available quality of recognition is high enough to address various use cases in the Digital Humanities. We argue that ASR will become a key technology for the documentation and analysis of audio-visual sources and identify an array of important questions that the DH community and cultural heritage stakeholders will have to address in the near future.

**Introduction**

Better availability of and access to digital technology has made it easier to create and share audio and video material in everyday life over the past decades. Future humanists will therefore face a mass of audio-visual sources. Automatic tools for finding, categorizing, indexing, analysing and archiving those sources will be required. We assume that Automatic Speech Recognition (ASR) will be among the key technologies to tackle these challenges.

The Digital Humanities (DH) is historically predominated by text-centric sources, but in recent years, we see more approaches to close the gap between Media Studies and DH. While automatic image and text recognition has been widely implemented (Bhargav et al. 2019; Lincoln et al. 2020; Wevers, Smits 2020; Lee et al. 2020; Taylor, Lauren 2019), we see significantly less projects focussing primarily on ASR-related tasks.

[a] One of these projects is the CLARIAH Media Suite using a self-created ASR model for Dutch (Van Noord et al. 2021).

Successful implementation would open an entire digital toolkit – from quantitative analysis to deep reading – for the analysis of the dominant medium of 20th and 21st century historical sources. ASR could be a key to rendering the content of mass communication archives – TV, radio and internet – searchable, it could allow automatic indexation and annotation and make an essential part of 20th and 21st centuries' societies' cultural heritage accessible to experts and public alike.

Section 1 introduces the speech model we created for German ASR and how it can be further used by the community. The following section 2 compares our model with commercially available services based on the corpus we built from historical recordings. Section 3 takes a close look on the transcripts that were created to find potentials and restrictions for the exploitation of ASR in Digital History, followed by a critical assessment of the limitations this technology currently faces in the last section, which further reflects our core findings.

**Building a state-of-the-art open-source ASR tool**

When the AI team of Meta (Facebook) released wav2vec in version 2.0 (Baevski et al.2020), they promised to provide an ASR framework which will be able to develop speech models which can be fine-tuned with less annotated training data than the currently most used frameworks Kaldi[b] and DeepSpeech[c] need for their trainings. The main idea of wav2vec2 is to split the training procedure into two separate parts: First, a speech model is produced with an enormous amount of unlabeled raw audio data in a self-supervised machine learning process. As the Meta team shows (Conneau et al. 2020: 4-6), this pretrained model does not necessarily have to be trained with the same language than the following fine-tuned model will be but can also be trained using multilingual corpora. The pretrained model then provides for fine-tuning with a speech model with relatively little annotated training data. wav2vec-based ASR-models have outperformed previous SOTA ASR-models even though they require significantly less fine-tuning with highly processed annotated audio data (Schneider et al. 2019).

---

[a] See also the I-Media-Cities project
[b] https://github.com/kaldi-asr/kaldi
[c] https://github.com/mozilla/DeepSpeech

For our study we chose the currently largest available pretrained speech model (Babu et al. 2021) which was trained with 436k hours of audio in 128 languages to fine-tune it with the German dataset of the CommonVoice project.[d] To keep the process simple, we used the Transformers Python modules provided by the HuggingFace AI community.[e] After pre-processing, the used training data sets consisted of about 225k unique spoken utterances along with their transcripts. On a single Tesla A100 GPU, the fine-tuning took roughly 80 hours. In a next step we added a n-gram language model to our speech model in order to improve the orthographical quality of the results. The language model was created with the kenfm framework.[f] We used the CommonVoice transcripts and the Europarl corpus as training data.[g] However, we did not see any quality improvements in the transcripts. We published all scripts used in this project on Github.[h] The final model is available in a HuggingFace repository.[i]

The final model can easily be used and tested with *.wav files at a sample rate of 44,1kHz by the following four lines of code after installing the Transformers and Torch libraries (*pip install transformers==4.17.0 torch==1.11.0*):

*from transformers import pipeline*

*pipe = pipeline(model="mfleck/wav2vec2-large-xls-r-300m-german-with-lm")*

*output = pipe("/path/to/file.wav",chunk_length_s=5, stride_length_s=1)*

*print(output["text"])*

Evaluating our wav2vec2 model against the German test dataset of CommonVoice achieves a word error rate of 8.8 percent. However, we do not consider that number very relevant, because the evaluation ran against the same dataset as the training, therefore the input audio signals during evaluation are likely to be very similar as the ones used for training. For this reason, we decided to proof our model against recordings which would be faced in practical use by scholars of Digital History.

---

[d] https://commonvoice.mozilla.org/de
[e] https://huggingface.co/docs/transformers/index
[f] https://github.com/kpu/kenlm
[g] https://opus.nlpl.eu/Europarl-v8.php
[h] https://github.com/MichaelFleck92/asr-wav2vec
[i] https://huggingface.co/mfleck/wav2vec2-large-xls-r-300m-german-with-lm

**Comparing results with commercially ASR systems**

We wanted to test the quality of transcripts for historical sources, so the publication dates of our selected recordings cover a period of time of roughly a century, from 1914 until the present day. Beside of this timely diversity, we covered different topics, numbers of speakers and dialects.[j] In total we transcribed 22 recordings manually, in order to build a body of reference, and compared them with the automatically generated transcripts of our own model.

To find out how our model performs against already existing solutions, we run our test corpus against four commercial ASR services. We decided to use three different cloud providers (Amazon Web Services (AWS), Google Cloud Platform (GCP) and Microsoft Azure) and the locally running dictating software Dragon Professional. The solutions provided by cloud providers represent classical black boxes to us as we can obtain only extremely limited knowledge on the software and hardware technologies used.

The following tables show the achieved word error rates and the word recognition rates based on a case-insensitive calculation.

| Word Error Rate (WER) | | | | | |
|---|---|---|---|---|---|
| Recording | AWS | GCP | Azure | Dragon | Wav2Vec2 |
| 1914 | 58,1 | 28,8 | 20,0 | 69,3 | 38,6 |
| 1916 | 65,0 | 87,8 | 29,7 | 92,0 | 40,7 |
| 1923 | 49,2 | 34,4 | 24,6 | 65,6 | 32,8 |
| 1929 | 21,3 | 20,1 | 14,2 | 25,9 | 21,3 |
| 1932 | 34,0 | 36,5 | 21,8 | 57,9 | 28,4 |
| 1938 | 20,6 | 28,2 | 14,4 | 29,1 | 20,0 |
| 1943 | 19,1 | 19,1 | 6,9 | 72,3 | 26,1 |
| 1947 | 42,8 | 70,5 | 43,5 | 82,1 | 52,6 |
| 1954 | 24,3 | 27,1 | 20,6 | 52,3 | 24,1 |
| 1958 | 14,9 | 10,6 | 7,4 | 42,6 | 28,7 |
| 1963 | 13,1 | 18,4 | 8,9 | 71,5 | 16,6 |
| 1964 | 16,9 | 16,9 | 12,1 | 24,6 | 19,5 |
| 1972 | 8,4 | 5,4 | 4,0 | 7,7 | 19,7 |
| 1976 | 48,6 | 56,4 | 39,3 | 71,7 | 59,6 |
| 1986 | 20,1 | 30,8 | 16,8 | 34,1 | 30,8 |
| 1988 | 13,6 | 13,6 | 9,5 | 22,8 | 17,9 |
| 1993 | 50,5 | 61,5 | 43,2 | 74,3 | 65,2 |
| 1999 | 19,5 | 25,6 | 12,7 | 25,1 | 14,4 |
| 2005 | 16,4 | 21,0 | 14,4 | 24,1 | 11,8 |
| 2006 | 13,7 | 28,3 | 17,0 | 25,9 | 19,1 |
| 2011 | 7,1 | 11,0 | 8,4 | 17,5 | 12,0 |
| 2018 | 8,2 | 10,6 | 3,8 | 29,3 | 10,3 |
| Average | 26,6 | 30,1 | 17,9 | 46,3 | 27,7 |
| Median | 19,8 | 26,3 | 14,4 | 38,3 | 22,7 |

---

[j] Austrian-German (recording 1947), Swiss-German (1976) and Bavarian-German (1993)

| Word Recognition Rate (WRR) | | | | | |
| --- | --- | --- | --- | --- | --- |
| Recording | AWS | GCP | Azure | Dragon | Wav2Vec2 |
| 1914 | 43,3 | 73,0 | 83,3 | 30,7 | 62,8 |
| 1916 | 35,4 | 12,2 | 71,9 | 8,0 | 61,6 |
| 1923 | 67,2 | 73,8 | 82,0 | 34,4 | 72,1 |
| 1929 | 84,9 | 83,3 | 88,3 | 74,9 | 80,8 |
| 1932 | 69,8 | 66,0 | 80,0 | 43,2 | 72,6 |
| 1938 | 83,5 | 74,4 | 88,2 | 72,4 | 82,1 |
| 1943 | 87,8 | 85,6 | 94,1 | 28,2 | 74,5 |
| 1947 | 61,8 | 30,2 | 59,3 | 17,9 | 48,4 |
| 1954 | 78,7 | 75,0 | 81,4 | 49,0 | 76,6 |
| 1958 | 91,5 | 91,5 | 95,7 | 58,5 | 75,5 |
| 1963 | 91,1 | 84,6 | 92,9 | 28,5 | 83,7 |
| 1964 | 86,2 | 87,0 | 90,7 | 77,4 | 83,1 |
| 1972 | 95,3 | 95,7 | 97,7 | 93,6 | 80,6 |
| 1976 | 60,4 | 44,7 | 69,0 | 28,6 | 41,6 |
| 1986 | 82,7 | 75,7 | 87,4 | 67,3 | 71,0 |
| 1988 | 91,7 | 89,8 | 92,5 | 80,3 | 83,8 |
| 1993 | 52,4 | 40,1 | 63,9 | 26,2 | 37,7 |
| 1999 | 85,1 | 78,8 | 90,0 | 78,5 | 86,8 |
| 2005 | 88,2 | 82,6 | 88,2 | 81,5 | 89,7 |
| 2006 | 94,1 | 77,9 | 93,8 | 80,6 | 84,4 |
| 2011 | 94,8 | 91,2 | 94,5 | 85,7 | 89,3 |
| 2018 | 94,4 | 90,6 | 96,8 | 72,4 | 90,0 |
| Average | 78,2 | 72,9 | 85,5 | 55,4 | 74,0 |
| Median | 85,0 | 78,3 | 88,3 | 62,9 | 78,6 |

**Results and discussion**

As we can see in the tables, our wav2vec2 model clearly outperforms the dictating software Dragon. It also achieves slightly better results than the cloud-based ASR service by Google and is close to the results from AWS. The Speech to Text service from Microsoft Azure is achieving the best results against our dataset.

When taking a closer look at the transcripts of our wav2vec2 model, we see that the framework has problems to split two words correctly, when they are pronounced quickly one after another or when dealing with dominant background noise. The recognition of single words is usually quite well, but the non-present hyphenation often generates a high word error rate although the word recognition is actually correct. In recording 2005 our model even achieved the best results. In this recording Jean Ziegler is speaking with a noticeable French accent, which gets handled better by the multilingual pretrained model than the solutions used by cloud service providers.

In our corpus we had three recordings with dialectical German. Those were recording 1947 (Austrian-German), recording 1974 (Swiss-German) and recording 1993 (Bavarian-German), which were all transcribed badly. Word error rates attained values around 40 to 60 percent each. This means that currently only standard German recordings can be a target of ASR projects. All cloud providers offer specific speech models for Austrian- and Swiss-German, which were used in the evaluation above. However, we also compared the results using their standard German models and we saw no significant differences in the outcome.

As we can see in the table, the newer a recording is, the better the quality of its transcript will be. The main reason is the worse audio quality of older recordings. Furthermore, trained statistical language models do not always fit older recordings, because language habits change significantly over decades.

In case that a corpus meets the conditions outlined above – good quality audio signal, standard language German – ASR achieves word recognition rates beyond 90 percent, which makes it a suitable tool to be used in Digital Humanities research, yet also in the indexation of audio-visual mass sources.

As we can see, open-source ASR tools can already achieve decent results, but if quality of recognition is the most important criteria in deciding which ASR tool to be used, we currently see no alternative to using cloud-based systems. Beside of the simple recognition of words, cloud providers already offer a wide range of complementary features. If the recognized words are correct, we see that the transcripts are orthographically and grammatically robust. Punctation and case sensitivity are properly used in most cases. Furthermore, ASR cloud services optionally deliver timestamps, which help to enrich a source. Not only what was said, but also when it was said, can be determined easily, which helps when it comes to automatised meta data enrichment. Especially when dealing with video files we see that as a valuable feature. Cloud-based ASR can also easily be integrated with speaker and intent recognition.

Transcription quality, however, might not be the most important feature when it comes to the functionalities many archives, libraries and similar stakeholders with regard to cultural heritage data and humanities are seeking: The automatic indexation, categorization and annotation of contents in large sets of audio and audio-visual data does not require outstanding transcription quality. Therefore, our model could already provide a solution for stakeholders that dispose of data sets with the features mentioned above – predominantly standard German, relatively little noise. We expect that recordings of TV and radio news for instance could benefit from our model. Further, future versions of our model will allow for better results, subsequently expanding the range of potential users.

**Critical assessment**

As has become clear from this article, ASR works particularly well under certain, very specific conditions, which have been outlined above: Standard German, good audio quality, sufficient signal-to-noise ratio. Its large-scale use bears a certain potential that sources that cannot be accessed via ASR will be further marginalized: Whereas our experience with similar technologies in the past shows that issues of quality can be tackled by technological solutions in many cases, we are less optimistic when it comes to the processing of sources that feature particularly specific challenges in terms of language. For once, German dialects run risk to disappear in studies building on large-scale digital analysis. The same phenomenon applies to the visibility and the representation of communities in DH research that make use of languages with a.) a limited number of speakers and b.) very little economic potential to those large-scale cloud solution providers whose software performed best in our research. During this study, we have also trained a model for the Slovenian language. Due to the lack of a sufficient amount of training data, the final model[k] only achieved a word error rate of 35 percent against the CommonVoice test dataset.

Whereas ASR offers new ways for the analysis, indexing and understanding of the bulk of languages for communities with enough annotated training material, minorities' cultural heritage is exposed to a significant risk of marginalization or even disappearance in certain branches of DH-based research. Our own experience in this area of research, built over the past two years, however, would let us assume that coming generations of models such as wav2vec, which implement an entirely different approach from its predecessors such as Mozilla DeepSpeech or Kaldi, will rather reduce this risk, as they display a remarkable capacity to deal with spoken language beyond what they were pretrained for.

Another major issue at stake here is a particular legal component: In order to benefit from the cloud-based services put to the test here, users have to accept terms of use that may interfere with stakeholders' interests, as cloud-operators reserve certain rights to exploit and reuse the data uploaded by customers. It is optimistic to state that this will affect only a minority of cases, on the contrary, this issue might turn out as a real blocker. Several options might contribute to a solution here: For once, customized solutions could be sought and agreed upon between cloud-operators and cultural heritage and research institutions, which appears to be more feasible for larger institutions than for smaller ones. From a global perspective, we expect that joint action of a majority of stakeholders in the field of cultural heritage and research institutions would be difficult to attain, yet would yield significantly better results. There is already a high degree of asymmetry in the relations between cloud-operators and data-owners. A technological solution could consist in the use of homomorphous encryption or transfer learning (Gentry 2009; Walch et al. 2022).

---

[k] https://huggingface.co/mfleck/wav2vec2-large-xls-r-300m-slowenian-with-lm

The usability of cloud-based solutions operated by major data companies in the context of scientific research remains a recurring issue, due to the fact that a.) there is a large number of parameters that cannot be controlled by researchers; b.) these systems do at least partially constitute black boxes, blocking the research community from breaking down and understanding significant parts of the process taking place and hindering the identification of potential biases; and c.) massive legal objections.

Powerful frameworks such as wav2vec2 could offer an alternative to working with commercial cloud-based operators, even though they currently do not attain their performance. We see considerable potential to further improve our model, particularly with regard to edge cases – specializing on historical recordings, recordings with a limited set of speakers (historical television and radio news shows), and especially speech featuring a high degree of dialect or strong accents. We do, however, not expect that we will be able to outperform Microsoft Azure anytime soon, particularly regarding standard speech situations. Past experiences in the context of DH show that research-led software development can hardly ever compete with the industry, yet the adaptation of open source SOTA solutions bears some promise, as we have seen in a large number of fields lately, from HTI to NLP.[i] Joint research and development projects between providers of mass audio-visual data corpora and research groups focusing on ASR could already implement powerful ASR functionality building on the available models, given they know how to design and set-up an efficient fine-tuning process. Such models could already master an array of different tasks, even though more specific requirements might remain currently out of range.

To sum up, more recent research into the subject of speech-to-text has delivered encouraging results when it comes to the adaptation and refinement of open-source ASR-solutions when combined with language models, which were built using NLP. Multidisciplinary cooperation across research areas has proved very promising here.

**Conclusion**

In this paper we analysed the performance of different ASR solutions based on the quality of transcripts of historical recordings in German. As deep learning algorithms are trained with everyday life audio data, the transcription of older recordings does not attain similarly high accuracy as the transcription of contemporary recordings produces, as the statistical language models do not match historical language patterns to a comparably high degree. However, as we point out, the biggest problem is the insufficient audio quality of ancient audio and video recordings. Another problem is the high number of different dialects and accents in German.

---

[i] Cf. Transkribus resp. the use of SpaCy in the context of digital editions.

The wav2vec2 ASR model we built for this study achieves moderate results in comparison with cloud-based tools by the big internet players. The main issue in developing open-source machine learning based tools with regard to their application in the cultural heritage context is that in an open-source project, the data used for training will get published as well. Big tech companies can use those data to improve their own proprietary tools, while they do not share their internal training data with the open-source community, which accounts for a major asymmetry here, particularly given the fact that large data companies can control the quality of the high-quality data their users and customers produce for them, which puts them at another advantage. Even if the wav2vec2 do not need that big amount of annotated than similar frameworks do, this major issue still persists for ASR tools. As long as the amount of training data is crucial for the quality of the final outcome, we see no change in this fundamental public-private issue.[m]

When using ASR projects in research projects with the goal to get the best possible recognition quality, we argue to use cloud-based solution if a) the corpus to be analysed is not extensible huge as cloud providers charge per usage and b) there are no legal restrictions in form of privacy issues with the corpus. Furthermore, cloud services provide a wide range of complimentary features without the need of resource intensive development beforehand making it very simple to use with a low entry barrier.

---

[m] Questions of scale, however, are a subject of large debates currently, and many data scientists do not believe that further scaling up of training data sets will be the key to better results. Cf. https://ai.googleblog.com/2022/04/pathways-language-model-palm-scaling-to.html

**Repositories and models**

https://github.com/kpu/kenlm

https://opus.nlpl.eu/Europarl-v8.php

https://www.imediacities.eu/

https://huggingface.co/mfleck/wav2vec2-large-xls-r-300m-german-with-lm